\documentclass[a4paper,english]{article}
\usepackage[T1]{fontenc}
\usepackage[latin9]{inputenc}
\usepackage{amsmath}
\usepackage{amssymb}
\usepackage{esint}

\makeatletter

\usepackage{a4wide}
\usepackage{booktabs}
\usepackage{hyperref}

\makeatother

\usepackage{babel}
\begin{document}

\title{On finite size corrections to the dispersion relations of giant magnon
and single spike on $\gamma$-deformed $T^{1,1}$}

\author{M.~Michalcik$^{\star}$\thanks{e-mail: kanister@hep.itp.tuwien.ac.at.}
and R.~C.~Rashkov$^{\star,\diamondsuit}$\thanks{e-mail: rash@hep.itp.tuwien.ac.at.}
\ \\
 \ \\
 $^{\star}$ Institute for Theoretical Physics, \\
 Vienna University of Technology,\\
 Wiedner Hauptstr. 8-10, 1040 Vienna, Austria, \ \\
 \ \\
 $^{\diamondsuit}$ Department of Physics, Sofia University \\
 5 J. Bourchier Blvd, 1164 Sofia, Bulgaria}
\maketitle
\begin{abstract}
In this paper we consider the finite size effects for the strings
in $\beta$-deformed $AdS_{5}\times T^{1,1}$ background. We analyze
the finite size corrections for the cases of giant magnon and single
spike string solution. The finite size corrections for the undeformed
case are straightforwardly obtained sending the deformation parameter
to zero.
\end{abstract}

\section{Introduction}

After the remarkable AdS/CFT conjecture of Maldacena \cite{Maldacena1998}
relating the string theory on $AdS_{5}\times S^{5}$ to the $\mathcal{N}=4$
super Yang-Mills the search for less supersymmetric cases  has started.
One possible way is to consider a stack of N D3-branes at the tip
of the conifold \cite{Klebanov:1998hh} resulting in the spacetime
being the product $AdS_{5}\times T^{1,1}$. The dispersion relations
on the string theory side plays an important role, since they correspond
to the large N limit of the anomalous dimension of particular gauge
theory operators .  Finding the spectrum  is however a complicated
task and one usually considers first states for which one of the string
charges is infinite. This simplification can be seen in the light-cone
gauge as a decompactifying limit, since one can rescale the worldsheet
coordinate, such that the theory is defined on a cylinder with circumference
proportional to the light-cone momentum (see \cite{Arutyunov2007a}
and references therein). In \cite{Hofman2006} it was shown that at
large $\lambda$ in the decompactifying limit a one-magnon excitation
with the finite worldsheet momentum can be identified with the classical
string sigma model. This string solution will have infinite momentum
and infinite energy, but the difference of the two will be finite
and equal to the energy of the worldsheet soliton \cite{Hofman2006,Arutyunov2007a}.
On the target space it corresponds to an open rigidly moving string
with the distance of the endpoints fixed  and proportional to the
worldsheet momentum. This solution constructed in the conformal gauge
was called \emph{the giant magnon} \cite{Hofman2006}. The situation
on the conifold was analyzed in \cite{Benvenuti2009}. 

One possible way of breaking the supersymmetry was found by Lunin
and Maldacena in \cite{Lunin2005}. They considered a marginally deformed
$\mathcal{N}=4$ super Yang-Mills and found its gravity dual. This
idea was further extended in \cite{Leigh1995,Frolov2005} and applied
to other backgrounds , see \cite{Dimov2009}.

Since all these considerations solve the theory on a plane, one can
ask what is the difference to the physical case of a cylinder worldsheet.
Such a modification can be significant, but for sufficiently large
 size of the cylinder one can use an asymptotic construction \cite{Arutyunov2007a}
and obtain the \emph{finite size corrections.} 

In this paper we extend our previous considerations \cite{Dimov2009}
computing the finite size corrections for the case of giant magnons
and single spikes on the $\gamma$-deformed $T^{1,1}$. Since the
deformation is continuous we obtain the finite size correction to
the undeformed case \cite{Benvenuti2009} as well, simply by setting
$\tilde{\gamma}=0$.

\section{A review of the case of infinite charges and finite size setup}

In this section we review the results of the case of infinite charges
\cite{Dimov2009}, reformulate some of the important steps of the
calculations to a convenient form and make a setup for the finite
size calculations. Setting $\tilde{\gamma}=0$ one can recover the
undeformed case $AdS_{5}\times T^{1,1}$ and the obtained results
translate to that case.

\subsection{Review of the infinite case}

The TsT transformations \cite{Lunin2005,Frolov2005,Catal-Ozer2006}
of the standard conifold metric results in the metric and the B-field
of the $\gamma$-deformed $T^{1,1}$ 
\begin{multline}
\frac{ds^{2}}{R^{2}}=ds_{AdS}^{2}+G\left(\frac{1}{6}\sum_{i=1}^{2}\left(G^{-1}d\theta_{i}^{2}+\sin^{2}\theta_{i}d\phi_{i}^{2}\right)+\right.\\
\left.\frac{1}{9}\left(d\psi+\cos\theta_{1}d\phi_{1}+\cos\theta_{2}d\phi_{2}\right)^{2}+\tilde{\gamma}\frac{\sin^{2}\theta_{1}\sin^{2}\theta_{2}}{324}d\psi^{2}\right),\label{eq:gamma_def_metric}
\end{multline}
\begin{multline}
\frac{B}{R^{2}}=\tilde{\gamma}G\left(\left(\frac{\sin^{2}\theta_{1}\sin^{2}\theta_{2}}{36}+\frac{\cos^{2}\theta_{1}\sin^{2}\theta_{2}+\cos^{2}\theta_{2}\sin^{2}\theta_{1}}{54}\right)d\phi_{1}\wedge d\phi_{2}+\right.\\
\left.\frac{\sin^{2}\theta_{1}\cos\theta_{2}}{54}d\phi_{1}\wedge d\psi-\frac{\cos\theta_{1}\sin^{2}\theta_{2}}{54}d\phi_{2}\wedge d\psi\right).\label{eq:gamma_def_B}
\end{multline}
The standard conifold metric is the one in (\ref{eq:gamma_def_metric})
with $\tilde{\gamma}=0$ and the conformal factor $G$ in (\ref{eq:gamma_def_metric}),(\ref{eq:gamma_def_B})
is 
\[
G^{-1}=1+\tilde{\gamma}\left(\frac{\cos^{2}\theta_{1}\sin^{2}\theta_{2}+\cos^{2}\theta_{2}\sin^{2}\theta_{1}}{54}+\frac{\sin^{2}\theta_{1}\sin^{2}\theta_{2}}{36}\right).
\]
Since the complete $T^{1,1}$ background amounts to a complicated
set of equations of motion we proceed with a consistent truncation
to the subspace defined by 
\[
\theta_{2}=0\;,\;\phi_{2}=\text{const.}
\]
The ansatz leading to a solitonic type solution is 
\begin{gather*}
t=\kappa\tau\:,\:\theta_{2}=0\:,\:\phi_{2}=\text{const.}\\
\Psi=\omega_{\psi}\tau-\psi\left(y\right)\:,\:\Phi=\omega_{\phi}\tau+\phi\left(y\right)\:,\:\theta=\theta\left(y\right),
\end{gather*}
where $y=c\sigma-d\tau$ and setting $R=1$ we restrict the metric
(\ref{eq:gamma_def_metric}) and the B-field (\ref{eq:gamma_def_B})
to 
\begin{align}
ds^{2}= & -dt^{2}+\frac{1}{6}d\theta^{2}+\frac{G}{6}\sin^{2}\theta d\phi^{2}+\frac{G}{9}\left(d\psi+\cos\theta d\phi\right)^{2}\nonumber \\
B= & G\tilde{\gamma}\frac{\sin^{2}\theta}{54}\label{eq:ansatz}\\
G^{-1}= & 1+\tilde{\gamma}\frac{\sin^{2}\theta}{54}.\nonumber 
\end{align}
The Lagrangian in the coordinates (\ref{eq:ansatz}) is of the form
\begin{multline*}
L=\dot{t}^{2}+\frac{c^{2}-d^{2}}{6}\theta'^{2}+\frac{G}{9}\left(1+\frac{\sin^{2}\theta}{2}\right)\left(-\left(\omega_{\phi}-d\phi'\right)^{2}+c^{2}\phi'^{2}\right)+\\
\frac{2}{9}G\cos\theta\left(-\left(\omega_{\psi}-d\psi'\right)\left(\omega_{\phi}-d\phi'\right)+c^{2}\psi'\phi'\right)+\frac{G}{9}\left(-\left(\omega_{\psi}-d\psi'\right)^{2}+c^{2}\psi'^{2}\right)+\\
2G\tilde{\gamma}\frac{\sin^{2}\theta}{54}c\left(\left(\omega_{\phi}-d\phi'\right)\psi'-\left(\omega_{\psi}-d\psi'\right)\phi'\right)
\end{multline*}
and the once integrated equations of motion are ($A_{\phi}$ and $A_{\psi}$
below are the integration constants) 
\begin{align*}
\left(c^{2}-d^{2}\right)\phi'+d\omega_{\phi}= & \frac{3\left(A_{\phi}-A_{\psi}\cos\theta\right)}{G\sin^{2}\theta}+\tilde{\gamma}\frac{c}{9}\left(\omega_{\psi}+\omega_{\phi}\cos\theta\right)\\
\left(c^{2}-d^{2}\right)\psi'+d\omega_{\psi}= & \frac{3\left(A_{\psi}-A_{\phi}\cos\theta\right)}{G\sin^{2}\theta}+\frac{3A_{\psi}}{2G}-\frac{\tilde{\gamma}}{9}c\left(\omega_{\psi}\cos\theta+\omega_{\phi}\left(1+\frac{\sin^{2}\theta}{2}\right)\right).
\end{align*}
The Virasoro constraints \cite{Dimov2009} read off 
\begin{multline*}
\frac{c^{2}+d^{2}}{6}\theta'^{2}+\frac{G}{9}\left(1+\frac{\sin^{2}\theta}{2}\right)\left(\left(\omega_{\phi}-d\phi'\right)^{2}+c^{2}\phi'^{2}\right)+2\frac{G}{9}\cos\theta\left(\left(\omega_{\psi}-d\psi'\right)\left(\omega_{\phi}-d\phi'\right)+\right.\\
\left.c^{2}\psi'\phi'\right)+\frac{G}{9}\left(\left(\omega_{\psi}-d\psi'\right)^{2}+c^{2}\psi'^{2}\right)=\kappa^{2},
\end{multline*}
 
\begin{multline*}
\frac{1}{6}\theta'^{2}+\frac{G}{9}\left(1+\frac{\sin^{2}\theta}{2}\right)\left(\phi'^{2}-\frac{\omega_{\phi}\phi'}{d}\right)+\frac{G}{9}\left(\psi'^{2}-\frac{\omega_{\psi}\psi'}{d}\right)+\\
\frac{G}{9}\cos\theta\left(2\phi'\psi'-\frac{\omega_{\psi}\phi'+\omega_{\phi}\psi'}{d}\right)=0
\end{multline*}
and the restriction on the parameters is just 
\begin{equation}
2d\kappa^{2}=A_{\phi}\omega_{\phi}+A_{\psi}\omega_{\psi}.\label{eq:vir_kappa}
\end{equation}
Using the second Virasoro constraint we obtain the equation for $\theta$
\begin{multline}
\theta'^{2}+\frac{1}{3\left(c^{2}-d^{2}\right)^{2}\sin^{2}\theta}\left(\frac{2c\omega_{\phi}-\tilde{\gamma}A_{\psi}}{2}\right)^{2}\cos^{4}\theta-\left(2c\omega_{\phi}-\tilde{\gamma}A_{\psi}\right)\left(2c\omega_{\psi}-\tilde{\gamma}A_{\phi}\right)\cos^{3}\theta-\\
\frac{1}{2}\left(\left(2c\omega_{\psi}+\tilde{\gamma}A_{\phi}\right)^{2}+2\left(2c\omega_{\phi}-\tilde{\gamma}A_{\psi}\right)^{2}+27A_{\psi}^{2}-\frac{18}{d}\left(c^{2}+d^{2}\right)\left(A_{\phi}\omega_{\phi}+A_{\psi}\omega_{\psi}\right)\right)\cos^{2}\theta+\\
\left(\left(2c\omega_{\phi}-\tilde{\gamma}A_{\psi}\right)\left(2c\omega_{\psi}+\tilde{\gamma}A_{\phi}\right)-54A_{\phi}A_{\psi}\right)\cos\theta+\frac{1}{2}\left(\left(2c\omega_{\psi}+\tilde{\gamma}A_{\phi}\right)^{2}+\frac{3}{2}\left(2c\omega_{\phi}-\tilde{\gamma}A_{\psi}\right)^{2}+\right.\\
\left.27\left(3A_{\psi}^{2}+2A_{\phi}^{2}\right)-\frac{18}{d}\left(c^{2}+d^{2}\right)\left(A_{\phi}\omega_{\phi}+A_{\psi}\omega_{\psi}\right)\right)=0.\label{eq:EoM_theta}
\end{multline}

\subsection{Finite size analysis setup}

In \cite{Dimov2009} we considered the case of infinite charges. In
contrast to this case the finite size corrections originate from the
expression 
\begin{equation}
\int_{0}^{w_{max}}d\sigma=\int_{0}^{w_{max}}\frac{dy}{c}=\frac{1}{c}\int_{u\left(0\right)}^{u\left(w_{max}\right)}\frac{du}{u'}\label{eq:ws_integrals}
\end{equation}
where $u=\cos^{2}\left(\frac{\theta\left(y\right)}{2}\right)$ and
the worldsheet coordinate $\sigma$ is in the range $\left\langle -w_{max},w_{max}\right\rangle $
with $w_{max}$ finite. Rewriting the equation of motion (\ref{eq:EoM_theta})
in $u\left(y\right)$ we obtain an equation accounting to the finite
size 
\begin{equation}
u'^{2}=-f^{2}\left(u-u_{0}\right)\left(u-u_{1}\right)\left(u-u_{2}\right)\left(u-u_{3}\right)=P_{4}\left(u\right).\label{eq:EoMu_root_form}
\end{equation}
The quartic polynomial $P_{4}\left(u\right)$ on the right hand side
of (\ref{eq:EoMu_root_form}) has in general four different roots
$u_{i}$. In order to have a well defined derivative $u'$ in the
worldsheet integral (\ref{eq:ws_integrals}) the integration has to
be taken in the positive region of the $P_{4}\left(u\right)$. We
are looking for certain string profiles, namely magnons and spikes,
which have one turning point. This condition in the $u$ coordinate
is simply $u\left(w_{max}\right)=u_{turn}=u_{j}$ i.e. the turning
point is one of the roots $u_{i}$. Our aim is to find finite size
corrections to the dispersion relations obtained in the infinite case,
therefore the limit $w_{max}\rightarrow\infty$ amounts to setting
the turning point to $u_{turn}=0$. Looking at the following relation
\[
y=\int_{u\left(0\right)}^{u}\frac{d\tilde{u}}{\tilde{u'}}=F\left[u\right]-F\left[u\left(0\right)\right]
\]
and the form of $u'$ (\ref{eq:EoMu_root_form}) it is clear that
demanding $y\rightarrow\infty$ for $u_{turn}\rightarrow0$ we actually
demand $F\left[u\right]\underset{u_{turn}\rightarrow0}{\longrightarrow}\infty$.
The latter is possible if there is another root $u_{k}$ that equals
$u_{turn}$ at $w_{max}=\infty$. In order to have finite $y$ for
any $u\neq0$ we demand  one of the roots, say $u_{0}$ to be $u_{0}=0$.
The polynomial $P_{4}\left(u\right)$ then changes to 
\[
P_{4}\left(u\right)=-f^{2}u\left(u-u_{1}\right)\left(u-u_{2}\right)\left(u-u_{3}\right),
\]
which results in the condition $A_{\phi}=-A_{\psi}$. Using the notations
of \cite{Dimov2009} 
\begin{equation}
B_{\phi}=2c\omega_{\phi}+\tilde{\gamma}A_{\phi}\quad,\quad B_{\psi}=2c\omega_{\psi}+\tilde{\gamma}A_{\phi}\label{eq:B_gen_def}
\end{equation}
we find $P_{4}\left(u\right)$ in the following form 
\begin{align}
P_{4}\left(u\right) & =-\frac{1}{3\left(c^{2}-d^{2}\right)^{2}}\left(B_{\phi}^{2}u^{4}-2B_{\phi}\left(B_{\phi}+B_{\psi}\right)u^{3}+\right.\nonumber \\
 & \frac{1}{2}\left(B_{\phi}^{2}-B_{\psi}^{2}+6B_{\phi}B_{\psi}-27A_{\phi}^{2}+\frac{18}{d}\left(c^{2}+d^{2}\right)A_{\phi}\left(\omega_{\phi}-\omega_{\psi}\right)\right)u^{2}+\nonumber \\
 & \frac{1}{2}\left(\left(B_{\psi}-B_{\phi}\right)^{2}+81A_{\phi}^{2}-\frac{18}{d}\left(c^{2}+d^{2}\right)A_{\phi}\left(\omega_{\phi}-\omega_{\psi}\right)\right)u.\label{eq:EoMu}
\end{align}
The prefactor $f^{2}$ is thus 
\[
f^{2}=\frac{B_{\phi}^{2}}{3\left(c^{2}-d^{2}\right)^{2}}.
\]
 The solution to the equation (\ref{eq:EoMu_root_form}) written
in the implicit form is 
\[
\int_{u3}^{u}\frac{d\tilde{u}}{-\left|f\right|\sqrt{\tilde{u}\left(\tilde{u}-u_{1}\right)\left(\tilde{u}-u_{2}\right)\left(u_{3}-\tilde{u}\right)}}=y,
\]
where we set $u_{2}$ to be the turning point  and $u_{3}>u_{2}>0$.
Analyzing the polynomial $P_{4}\left(u\right)$ one finds that $u_{1}<0$.
In order to work with positive parameters $u_{i}$, it is convenient
to define $u_{11}=-u_{1}$.

\section{Finite size corrections}

In this section we proceed with the conserved charges. Referring to
\cite{Dimov2009} for further details, we start with the relations
for conserved charges in the finite case 
\[
J_{i}=2\int_{0}^{w_{max}}\frac{dy}{c}P_{i}=\frac{2}{c}\int_{u_{3}}^{u_{2}}\frac{du}{u'}P_{i}.
\]
Using the notations 
\begin{equation}
I_{1}=\int_{u_{3}}^{u_{2}}\frac{du}{u'}\,;\, I_{2}=\int_{u_{3}}^{u_{2}}u\frac{du}{u'}\,;\, I_{3}=\int_{u_{3}}^{u_{2}}u^{2}\frac{du}{u'}\,;\, I_{4}=\int_{u_{3}}^{u_{2}}\frac{1}{1-u}\frac{du}{u'}\label{eq:Integrals_def}
\end{equation}
we write the general expression for conserved charges 
\begin{align}
cE= & 2T\kappa I_{1}\nonumber \\
9c\left(c^{2}-d^{2}\right)J_{\psi}= & T\left(9dA_{\phi}-c\left(B_{\phi}-B_{\psi}\right)\right)I_{1}+2cTB_{\phi}I_{2}\nonumber \\
9c\left(c^{2}-d^{2}\right)J_{\phi}= & T\left(c\left(B_{\phi}-B_{\psi}\right)-9dA_{\phi}\right)I_{1}+2cT\left(B_{\phi}+B_{\psi}\right)I_{2}-2cTB_{\phi}I_{3}\label{eq:charsI}\\
9c\left(c^{2}-d^{2}\right)\Delta= & \frac{9}{2}\left(3cA_{\phi}-d\left(B_{\phi}-B_{\psi}\right)\right)I_{1}+\tilde{\gamma}c\left(\left(2B_{\phi}+B_{\psi}\right)I_{2}-B_{\phi}I_{3}\right)+27cA_{\phi}I_{4}.\nonumber 
\end{align}
The integrals $I_{i}$ are expressed in terms of complete elliptic
integrals (\ref{eq:explicit_integrals}) and thus the above set of
equations (\ref{eq:charsI}) becomes  transcendental one.  From
the Virasoro constraints (\ref{eq:vir_kappa}) we have 
\[
\kappa=\sqrt{\frac{A_{\phi}\left(\omega_{\phi}-\omega_{\psi}\right)}{2d}},
\]
which results in 
\begin{equation}
I_{1}=\frac{c\sqrt{d}E}{T\sqrt{2A_{\phi}\left(\omega_{\phi}-\omega_{\psi}\right)}}.\label{eq:I1msolG}
\end{equation}
In order to proceed we assume that our turning point $u_{2}$ is very
small i.e. $u_{2}\mapsto\epsilon\rightarrow0$ which is equivalent
to having large $w_{max}$ and so large, but finite charges. In the
rest of this paper we expand our calculations in $\epsilon\equiv u_{2}$.
The turning point condition $u'\left(\epsilon\right)=0+O\left(\epsilon\right)^{2}$
gives again two relations for the parameter $A_{\phi}$ defining the
magnon and the spike solution 
\begin{equation}
A_{\phi}\begin{cases}
\begin{gathered}\frac{2}{9}d\left(\omega_{\phi}-\omega_{\psi}\right)+\frac{1}{243\left(c^{2}-d^{2}\right)\left(\omega_{\phi}-\omega_{\psi}\right)}\\
4d\left(\left(27c^{2}+12dc\tilde{\gamma}+d^{2}\left(\tilde{\gamma}^{2}+9\right)\right)\omega_{\phi}^{2}+d\left(d\left(\tilde{\gamma}^{2}+9\right)-6c\tilde{\gamma}\right)\omega_{\psi}^{2}+\right.\\
\left.2\left(27c^{2}-3dc\tilde{\gamma}-d^{2}\left(\tilde{\gamma}^{2}+9\right)\right)\omega_{\psi}\omega_{\phi}\right)\epsilon+O\left(\epsilon^{2}\right)
\end{gathered}
 & \text{magnon}\\
\begin{gathered}\frac{2c^{2}}{9d}\left(\omega_{\phi}-\omega_{\psi}\right)+\frac{1}{243d\left(c^{2}-d^{2}\right)\left(\omega_{\phi}-\omega_{\psi}\right)}\\
4c^{2}\left(\left(27d^{2}+12dc\tilde{\gamma}+c^{2}\left(\tilde{\gamma}^{2}+9\right)\right)\omega_{\phi}^{2}+c\left(c\left(\tilde{\gamma}^{2}+9\right)-6d\tilde{\gamma}\right)\omega_{\psi}^{2}+\right.\\
\left.2\left(27d^{2}-3dc\tilde{\gamma}-d^{2}\left(\tilde{\gamma}^{2}+9\right)\right)\omega_{\psi}\omega_{\phi}\right)\epsilon+O\left(\epsilon^{2}\right)
\end{gathered}
 & \text{spike}
\end{cases}\label{eq:A_phi_sol}
\end{equation}

Since $A_{\phi}$ receives $\epsilon$-correction so does $B_{\phi},B_{\psi}$,
therefore for convenience we write them as 
\begin{equation}
B_{\phi,\psi}=B0_{\phi,\psi}+\epsilon B1_{\phi,\psi}+O\left(\epsilon\right)^{2}.\label{eq:B_expansion}
\end{equation}
Using the above form of parameters $B_{\phi,\psi}$ (\ref{eq:B_expansion})
and the expansions of the elliptic integrals (\ref{eq:ellexpansion})
we find the following expressions for the integrals $I_{i}$ :
\begin{align}
I_{1}= & \frac{\sqrt{3}\left(c^{2}-d^{2}\right)}{B0_{\phi}\sqrt{u_{11}u_{3}}}\left(\log\left(\frac{16u_{11}u_{3}}{\left(u_{11}+u_{3}\right)\epsilon}\right)+\frac{1}{4u_{11}u_{3}}\right.\nonumber \\
 & \left(\left(u_{11}-u_{3}-4\frac{B1_{\phi}}{B0_{\phi}}u_{11}u_{3}\right)\log\left(\frac{16u_{11}u_{3}}{\left(u_{11}+u_{3}\right)\epsilon}\right)-2\left(u_{11}-u_{3}\right)\right)\epsilon+O\left(\epsilon\right)^{2}\nonumber \\
I_{2}= & \frac{\sqrt{3}\left(c^{2}-d^{2}\right)}{B0_{\phi}}\left(2\arctan\left(\sqrt{\frac{u_{3}}{u_{11}}}\right)+\right.\nonumber \\
 & \left.\left(\frac{B0_{\phi}}{2\sqrt{3}\left(c^{2}-d^{2}\right)}I_{1}-\frac{1}{2\sqrt{u_{11}u_{3}}}-2\frac{B1_{\phi}}{B0_{\phi}}\arctan\left(\sqrt{\frac{u_{3}}{u_{11}}}\right)\right)\epsilon\right)+O\left(\epsilon\right)^{2}\nonumber \\
I_{3}= & \frac{\sqrt{3}\left(c^{2}-d^{2}\right)}{B0_{\phi}}\left(\sqrt{u_{11}u_{3}}+\left(u_{3}-u_{11}\right)\arctan\left(\sqrt{\frac{u_{3}}{u_{11}}}\right)+\right.\label{eq:Integrals}\\
 & \left.\left(-\frac{B1_{\phi}}{B0_{\phi}}\sqrt{u_{11}u_{3}}+\left(1+\frac{B1_{\phi}}{B0_{\phi}}\left(u_{11}-u_{3}\right)\right)\arctan\left(\sqrt{\frac{u_{3}}{u_{11}}}\right)\right)\epsilon\right)+O\left(\epsilon\right)^{2}\nonumber \\
I_{4}= & I_{1}+\frac{6\left(c^{2}-d^{2}\right)}{\sqrt{3}B0_{\phi}\sqrt{\left(1+u_{11}\right)\left(1-u_{3}\right)}}\arctan\left(\sqrt{\frac{\left(1+u_{11}\right)u_{3}}{u_{11}\left(1-u_{3}\right)}}\right)+\left(\frac{1}{2}I_{1}+\right.\nonumber \\
 & \left.\frac{\sqrt{3}\left(c^{2}-d^{2}\right)}{2B0_{\phi}\sqrt{u_{11}u_{3}}}\left(-1+2\left(1-2\frac{B1_{\phi}}{B0_{\phi}}\right)\sqrt{\frac{u_{11}u_{3}}{\left(1+u_{11}\right)\left(1-u_{3}\right)}}\right)\arctan\left(\sqrt{\frac{\left(1+u_{11}\right)u_{3}}{u_{11}\left(1-u_{3}\right)}}\right)\right)\epsilon+O\left(\epsilon\right)^{2}\nonumber 
\end{align}
One see that the first integral $I_{1}$ is divergent for $\epsilon\rightarrow0$
and behaves as  $\left.I_{1}\right|_{\epsilon\rightarrow0}\sim-\log\left(\epsilon\right)$.
Using this fact we get from the integral $I_{1}$ (\ref{eq:Integrals})
\begin{equation}
\frac{I_{1}\sqrt{u0_{11}u0_{3}}B0_{\phi}}{\sqrt{3}\left(c^{2}-d^{2}\right)}+\log\left(\frac{\left(u0_{11}+u0_{3}\right)}{16u0_{11}u0_{3}}\epsilon\right)+O\left(\epsilon^{1}\right)=0.\label{eq:epsilon_eq}
\end{equation}
Substituting $I_{1}$ from (\ref{eq:I1msolG}) and solving (\ref{eq:epsilon_eq})
we find 
\begin{equation}
\epsilon\approx\frac{16e^{\mathcal{E}X}u0_{11}u0_{3}}{u0_{11}+u0_{3}}\;,\; X=-\frac{B_{\phi}c\sqrt{d}\sqrt{u_{11}u_{3}}}{\sqrt{6}\left(c^{2}-d^{2}\right)\sqrt{A_{\phi}\left(\omega_{\phi}-\omega_{\psi}\right)}}\;,\;\mathcal{E}=E/T.\label{eq:epsilon_relation}
\end{equation}
We can now return to the calculation of $u_{11},u_{3}$ from (\ref{eq:EoMu}).
In order to expand in the $\epsilon$ parameter first we write 
\begin{equation}
u_{11,3}=u0_{11,3}+\epsilon u1_{11,3}.\label{eq:u_exp}
\end{equation}
The turning point condition simplifies $P_{4}\left(u\right)=0$ (\ref{eq:EoMu})
to the form
\begin{multline}
u\left(27A0_{\phi}^{2}+(u-1)B0_{\phi}\left(uB0_{\phi}-B0_{\phi}-2B0_{\psi}\right)\right)+\epsilon\left(54uA0_{\phi}A1_{\phi}+27\left(u-1\right)A0_{\phi}^{2}+\left(u-1\right)\right.\\
\left.\left(2u^{2}B0_{\phi}B1_{\phi}-2u\left(B0_{\psi}B1_{\phi}+B0_{\phi}\left(B1_{\psi}+B1_{\phi}\right)\right)+B0_{\phi}\left(B0_{\phi}+2B0_{\psi}\right)\right)\right)+O\left(\epsilon^{2}\right)=0,\label{eq:EoMuG}
\end{multline}
where we expanded the parameter $A_{\phi}$ in $\epsilon$ as $A_{\phi}=A0_{\phi}+\epsilon A1_{\phi}$.
The solution to (\ref{eq:EoMuG}) at the zeroth order in $\epsilon$
, which we call $u0_{11,3}$ , coincides with the infinite case results
in \cite{Dimov2009}. Thus we have 
\begin{equation}
u_{3}=1+\frac{B0_{\psi}}{B0_{\phi}}+\frac{\sqrt{B0_{\psi}{}^{2}-27A0_{\phi}^{2}}}{B0_{\phi}}+\epsilon u1_{3}\:,\:-u_{1}\equiv u_{11}=-1-\frac{B0_{\psi}}{B0_{\phi}}+\frac{\sqrt{B0_{\psi}{}^{2}-27A0_{\phi}^{2}}}{B0_{\phi}}+\epsilon u1_{11}\label{eq:roots_0}
\end{equation}
with 
\begin{align}
u1_{3}= & \frac{-1}{2B0_{\phi}{}^{2}\sqrt{B0_{\psi}{}^{2}-27A0_{\phi}^{2}}}\left(27A0_{\phi}^{2}\left(B0_{\phi}-2B1_{\phi}\right)+54A0_{\phi}B0_{\phi}A1_{\phi}+\right.\nonumber \\
 & \left.\left(2B0_{\psi}B1_{\phi}-2B0_{\phi}B1_{\psi}+B0_{\phi}{}^{2}\right)\left(\sqrt{B0_{\psi}{}^{2}-27A0_{\phi}^{2}}+B0_{\psi}\right)\right)\nonumber \\
u1_{11}= & \frac{-1}{2B0_{\phi}{}^{2}\sqrt{B0_{\psi}{}^{2}-27A0_{\phi}^{2}}}\left(27A0_{\phi}^{2}\left(B0_{\phi}-2B1_{\phi}\right)+54A0_{\phi}B0_{\phi}A1_{\phi}-\right.\label{eq:roots_epsilon}\\
 & \left.\left(2B0_{\psi}B1_{\phi}-2B0_{\phi}B1_{\psi}+B0_{\phi}{}^{2}\right)\left(\sqrt{B0_{\psi}{}^{2}-27A0_{\phi}^{2}}-B0_{\psi}\right)\right)\nonumber 
\end{align}
 In the further analysis we consider the different cases of giant
magnon and single spike.

\subsection{Giant magnon}

We start with the value of $A_{\phi}$ (\ref{eq:A_phi_sol}) specifying
the magnon solution. In order to find the dispersion relation we will
substitute the integrals $I_{i}$ from (\ref{eq:Integrals}),(\ref{eq:I1msolG})
to the charges (\ref{eq:charsI}), use (\ref{eq:u_exp}) as well as
the relations  
\begin{align*}
Bm0_{\psi}= & -\frac{1}{2}\left(u0_{11}-u0_{3}+2\right)Bm0_{\phi}\\
Bm1_{\psi}= & \frac{1}{2}\left(\left(-u0_{11}+u0_{3}-2\right)Bm1_{\phi}+\left(-u1_{11}+u1_{3}+1\right)Bm0_{\phi}\right)\\
\frac{Bm1_{\phi}}{Bm0_{\phi}}= & \frac{d\gamma\left(d^{2}\left(u0_{11}-u0_{3}+4\right){}^{2}-12c^{2}\left(u0_{11}-u0_{3}+1\right)\right)}{27c\left(c^{2}-d^{2}\right)\left(u0_{11}-u0_{3}+4\right)}+O\left(\epsilon\right).
\end{align*}
The parameters $BmI_{\alpha}$ above are just $B0_{\phi,\psi}\:,\: B1_{\phi,\psi}$
from (\ref{eq:B_expansion}) with the value of $A_{\phi}$ defining
the magnon case, i.e. $BmI_{\alpha}=\left.BI_{\alpha}\right|_{A_{\phi}=A_{\phi magnon}}\:,\: I\in\left\{ 0,1\right\} \,,\,\alpha\in\left\{ \phi,\psi\right\} $.
Thus we obtain relations for $J_{\phi}\,,\, J_{\psi}$ accounting
for finite size effects 
\begin{multline}
\frac{4}{\sqrt{3}}\arctan\left(\sqrt{\frac{u0_{3}}{u0_{11}}}\right)=\mathcal{E}+3\mathcal{J}_{\psi}+\epsilon\left(\frac{2\left(u_{1}{}_{11}u0_{3}-u_{1}{}_{3}u0_{11}\right)}{\sqrt{3}\sqrt{u0_{11}u0_{3}}\left(u0_{11}+u0_{3}\right)}+\frac{1}{\sqrt{3}\sqrt{u0_{11}u0_{3}}}+\right.\\
\left.\frac{d^{2}\mathcal{E}\left(d^{2}\left(u0_{11}{}^{2}+\left(2-14u0_{3}\right)u0_{11}+u0_{3}{}^{2}-2u0_{3}-8\right)-6c^{2}\left(u0_{11}-u0_{3}-2\right)\right)}{36\left(c^{2}-d^{2}\right)^{2}\left(u0_{11}+1\right)\left(u0_{3}-1\right)}\right)+O\left(\epsilon^{2}\right)\label{eq:J_psi_eq}
\end{multline}
and 
\begin{multline}
\frac{2\sqrt{u0_{11}u0_{3}}}{\sqrt{3}}=\left(\mathcal{E}-3\mathcal{J}_{\phi}\right)+\frac{\epsilon}{36\left((c^{2}-d^{2})^{2}u0_{11}\left(u0_{11}+1\right)\left(u0_{3}-1\right)u0_{3}\right)}\left(\right.\\
\mathcal{E}u0_{11}u0_{3}d^{2}\left(3c^{2}\left(\left(u0_{11}-u0_{3}\right){}^{2}-4\right)-2d^{2}\left(u0_{11}{}^{2}+\left(4u0_{3}+2\right)u0_{11}+\left(u0_{3}-4\right)\left(u0_{3}+2\right)\right)\right)-\\
\left.6\sqrt{3}(c^{2}-d^{2})^{2}\left(u0_{11}+1\right)\left(u0_{3}-1\right)\sqrt{u0_{11}u0_{3}}\left(2u_{1}{}_{11}u0_{3}+\left(2u_{1}{}_{3}-1\right)u0_{11}+u0_{3}\right)\right)+O\left(\epsilon^{2}\right).\label{eq:J_phi_eq}
\end{multline}
Above we used the charge densities $\mathcal{J}_{i}=J_{i}/T$ instead
of charges. From the relations (\ref{eq:J_psi_eq}),(\ref{eq:J_phi_eq})
we find at the zeroth order approximation in $\epsilon$
\begin{equation}
u0_{11}=\frac{\sqrt{3}}{2}\left(\mathcal{E}-3\mathcal{J}_{\phi}\right)\cot\left(\frac{\sqrt{3}}{4}\left(\mathcal{E}+3\mathcal{J}_{\psi}\right)\right)\:,\: u0_{3}=\frac{\sqrt{3}}{2}\left(\mathcal{E}-3\mathcal{J}_{\phi}\right)\tan\left(\frac{\sqrt{3}}{4}\left(\mathcal{E}+3\mathcal{J}_{\psi}\right)\right).\label{eq:u0_sol_mag}
\end{equation}
 Substituting $u0_{11,3}$ (\ref{eq:u0_sol_mag}) to the angle difference
equation we recover the infinite case dispersion relation at the zeroth
order. At the first order we find the corrections $u1_{11,3}$ which
can be found in the appendix (\ref{eq:u1_corr_magnon}). The angle
difference equation written in a concise form is 
\begin{equation}
2\arctan\left(\sqrt{\frac{\left(u0_{11}+1\right)u0_{3}}{u0_{11}-u0_{11}u0_{3}}}\right)=\Delta-\frac{1}{2}\left(\mathcal{J}_{\phi}+\mathcal{J}_{\psi}\right)\gamma+\frac{27\Delta_{\epsilon}}{c\left(c^{2}-d^{2}\right)}\epsilon+O\left(\epsilon^{2}\right),\label{eq:disp_m_1}
\end{equation}
where $\Delta_{\epsilon}=\mathcal{E}\Delta_{\mathcal{E}}+\Delta_{1}$
and the parts $\Delta_{\mathcal{E}},\Delta_{1}$ are given in the
appendix (see \ref{eq:Delta_mag}). Since we assume $\epsilon\ll1$
and thus the energy density $\mathcal{E}$ being very large, we see
that the coefficient $\Delta_{\mathcal{E}}$ produces the leading
part of the correction and the $\Delta_{1}$ the subleading one. Expressing
$\Delta-\frac{1}{2}\left(\mathcal{J}_{\phi}+\mathcal{J}_{\psi}\right)\gamma$
from (\ref{eq:disp_m_1}) in a concise form \cite{Dimov2009} we obtain
\begin{multline}
\cos\left(\Delta-\frac{1}{2}\left(\mathcal{J}_{\phi}+\mathcal{J}_{\psi}\right)\gamma\right)=\frac{u0_{11}-u0_{3}-2u0_{3}u0_{11}}{u0_{11}+u0_{3}}+\\
\frac{108\Delta_{\epsilon}u0_{11}u0_{3}\left(1+u0_{11}\right)\left(1-u0_{3}\right)\epsilon}{c\left(c^{2}-d^{2}\right)\left(u0_{11}+u0_{3}\right){}^{2}\sin\left(\Delta-\frac{1}{2}\left(\mathcal{J}_{\phi}+\mathcal{J}_{\psi}\right)\gamma\right)}+O\left(\epsilon^{2}\right).\label{eq:disp_magnon_u1}
\end{multline}
Restricted to the first term it is just the infinite case dispersion
relation. 

In order to find the dispersion relation in terms of conserved charges
we must substitute for the $\epsilon$ parameter from (\ref{eq:epsilon_relation}).
Using the convenient notations $R_{\phi}=\frac{\sqrt{3}}{2}\left(\mathcal{E}-3\mathcal{J}_{\phi}\right)\:,\: R_{\psi}=\frac{\sqrt{3}}{2}\left(\mathcal{E}+3\mathcal{J}_{\psi}\right)\:,\: R_{\Delta}=\Delta-\frac{\gamma}{2}\left(\mathcal{J}_{\phi}+\mathcal{J}_{\psi}\right)$
we obtain the giant magnon dispersion relation (\ref{eq:disp_magnon_u1})
as 
\begin{align}
\cos\left(R_{\Delta}\right) & =\left(\cos\left(R_{\psi}\right)-R_{\phi}\sin\left(R_{\psi}\right)\right)\left(1+C_{1}\mathcal{E}e^{\mathcal{E}X}+D_{1}e^{\mathcal{E}X}+O\left(e^{\mathcal{E}X}\right)^{2}\right).\label{eq:disp_f_mag}
\end{align}
The finite size correction consists of two parts $C_{1}\mathcal{E}e^{\mathcal{E}X}$
and $D_{1}e^{\mathcal{E}X}$ where $C_{1},D_{1}$ are lengthy expressions
of $R_{\phi},R_{\psi},R_{\Delta}$, that can be found in the appendix
(\ref{eq:magnon_C1}),(\ref{eq:magnon_D1}). The exponent $X$ in
(\ref{eq:disp_f_mag}) is of the form 
\begin{align*}
X=- & \frac{2\sqrt{3}R_{\phi}\sin\left(R_{\psi}\right)\left(R_{\phi}\cos\left(2R_{\psi}\right)+2\sin\left(R_{\psi}\right)\right)}{1+4R_{\phi}^{2}-\left(1+2R_{\phi}^{2}\right)\cos\left(2R_{\psi}\right)-2R_{\phi}\sin\left(2R_{\psi}\right)}<0,
\end{align*}
or equivalently 
\[
X=-\frac{3c^{2}\left(\mathcal{E}-3\mathcal{J}_{\phi}\right)}{\left(c^{2}-d^{2}\right)\left(4+\sqrt{3}\left(\mathcal{E}-3\mathcal{J}_{\phi}\right)\cot\left(\frac{\sqrt{3}}{2}\left(\mathcal{E}+3\mathcal{J}_{\psi}\right)\right)\right)},
\]
which shows exponential suppression.

\subsection{Single spike}

For the case of a single spike defined by the value of $A_{\phi}$
(\ref{eq:A_phi_sol}) we use the same procedure as in the magnon case.
In order to find the dispersion relation including the finite size
correction we substitute the integrals $I_{i}$ from (\ref{eq:Integrals}),(\ref{eq:I1msolG})
to the relations for conserved charges (\ref{eq:charsI}) and use
the relations :

\begin{align*}
Bs0_{\psi}= & -\frac{1}{2}\left(u0_{11}-u0_{3}+2\right)Bs0_{\phi}\\
Bs1_{\psi}= & \frac{1}{2}\left(\left(-u0_{11}+u0_{3}-2\right)Bs1_{\phi}+\left(-u1_{11}+u1_{3}+1\right)Bs0_{\phi}\right)\\
\frac{Bs1_{\phi}}{Bs0_{\phi}}= & \frac{c\gamma\left(c^{2}\left(u0_{11}-u0_{3}+4\right){}^{2}-12d^{2}\left(u0_{11}-u0_{3}+1\right)\right)}{27d\left(c^{2}-d^{2}\right)\left(u0_{11}-u0_{3}+4\right)}+O\left(\epsilon\right).
\end{align*}
Here $BsI_{\alpha}=\left.BI_{\alpha}\right|_{A_{\phi}=A_{\phi spike}}\:,\: I\in\left\{ 0,1\right\} \,,\,\alpha\in\left\{ \phi,\psi\right\} $
from (\ref{eq:B_expansion}) and (\ref{eq:A_phi_sol}). Introducing
the charge densities $\mathcal{J}_{i}=J_{i}/T$ we obtain the following
expressions for the conserved charges $J_{\phi}\,,\, J_{\psi}$ (\ref{eq:charsI})
: 
\begin{multline}
\frac{4}{\sqrt{3}}\arctan\left(\sqrt{\frac{u0_{3}}{u0_{11}}}\right)=3\mathcal{J}_{\psi}+\epsilon\left(\frac{c^{3}\mathcal{E}\left(\left(u0_{11}-u0_{3}+1\right)\left(u0_{11}-u0_{3}+4\right)c^{2}+9d^{2}\left(u0_{3}-u0_{11}\right)\right)}{18d\left(c^{2}-d^{2}\right)^{2}\left(u0_{11}+1\right)\left(1-u0_{3}\right)}+\right.\\
\left.\frac{\sqrt{u0_{11}u0_{3}}\left(-2u1_{3}u0_{11}+u0_{11}+2u1_{11}u0_{3}+u0_{3}\right)}{\sqrt{3}u0_{11}u0_{3}\left(u0_{11}+u0_{3}\right)}\right)+O\left(\epsilon^{2}\right)\label{eq:J_psi_eq_spike}
\end{multline}
and 
\begin{multline}
\mathcal{J}_{\phi}=-\frac{2\sqrt{u0_{11}u0_{3}}}{3\sqrt{3}}-\frac{1}{108}\epsilon\left(\frac{6\sqrt{3}\left(2u1_{11}u0_{3}+u0_{3}+u0_{11}\left(2u1_{3}-1\right)\right)}{\sqrt{u0_{11}u0_{3}}}+\right.\\
\frac{c^{3}\mathcal{E}}{\left(c^{2}-d^{2}\right)^{2}\left(u0_{11}+1\right)\left(1-u0_{3}\right)}\left(2\sqrt{3}c\left(u0_{11}-u0_{3}-8\right)\sqrt{\left(u0_{11}+1\right)\left(1-u0_{3}\right)}\right.\\
\left.\left.-3d\left(u0_{11}{}^{2}-2\left(u0_{3}+2\right)u0_{11}+u0_{3}{}^{2}+4u0_{3}-8\right)\right)\right)+O\left(\epsilon^{2}\right)\label{eq:J_phi_eq_spike}
\end{multline}

Again, at the zeroth order in $\epsilon$ we find for the parameters
$u0_{11,3}$
\begin{equation}
u0_{11}=-\frac{3}{2}\sqrt{3}\mathcal{J}_{\phi}\cot\left(\frac{3\sqrt{3}}{4}\mathcal{J}_{\psi}\right)\quad,\quad u0_{3}=-\frac{3}{2}\sqrt{3}\mathcal{J}_{\phi}\tan\left(\frac{3\sqrt{3}}{4}\mathcal{J}_{\psi}\right)\label{eq:u0_spike}
\end{equation}
 which substituted to the remaining conserved quantity, the angle
difference $\Delta$, gives the infinite case dispersion relation.
From the equations (\ref{eq:J_psi_eq_spike}),(\ref{eq:J_phi_eq_spike})
we calculate the corrections $u1_{11,3}$ , which are given in the
appendix (see \ref{eq:u1_corr_spike}). Using (\ref{eq:u0_spike}),(\ref{eq:u1_corr_spike})
we find 
\begin{equation}
2\arctan\left(\sqrt{\frac{\left(u0_{11}+1\right)u0_{3}}{u0_{11}-u0_{11}u0_{3}}}\right)=\Delta-\frac{3}{2}\mathcal{E}-\frac{1}{2}\left(\mathcal{J}_{\phi}+\mathcal{J}_{\psi}\right)\gamma+\frac{27\Delta_{\epsilon}}{c\left(c^{2}-d^{2}\right)}\epsilon+O\left(\epsilon^{2}\right),\label{eq:disp_s_1}
\end{equation}

with $\Delta_{\epsilon}=\mathcal{E}\Delta_{\mathcal{E}}+\Delta_{1}$.
The leading $\Delta_{\mathcal{E}}$ and the subleading $\Delta_{1}$
part of $\Delta_{\epsilon}$ are in the appendix (\ref{eq:Delta_spike}).
The other convenient form of (\ref{eq:disp_s_1}) reads off 
\begin{multline}
\cos\left(\Delta-\frac{3}{2}\mathcal{E}-\frac{1}{2}\left(\mathcal{J}_{\phi}+\mathcal{J}_{\psi}\right)\gamma\right)=\frac{u0_{11}-u0_{3}-2u0_{3}u0_{11}}{u0_{11}+u0_{3}}+\\
\frac{108\Delta_{\epsilon}u0_{11}u0_{3}\left(1+u0_{11}\right)\left(1-u0_{3}\right)\epsilon}{c\left(c^{2}-d^{2}\right)\left(u0_{11}+u0_{3}\right){}^{2}\sin\left(\Delta-\frac{3}{2}\mathcal{E}-\frac{1}{2}\left(\mathcal{J}_{\phi}+\mathcal{J}_{\psi}\right)\gamma\right)}+O\left(\epsilon^{2}\right),\label{eq:disp_spike_u}
\end{multline}
with its zero-order being just the infinite case dispersion relation
for single spike \cite{Dimov2009}.

In order to find the explicit form of the corrections we substitute
for $\epsilon$ from (\ref{eq:epsilon_relation}) and for $u0_{11,3}$
from (\ref{eq:u0_spike}) and obtain 
\begin{equation}
\cos\Delta\delta=\frac{3\sqrt{3}}{2}\left(\mathcal{J}_{\phi}\sin\left(R_{\psi}\right)+\frac{2\cos\left(R_{\psi}\right)}{3\sqrt{3}}\right)\left(1+C_{1}\mathcal{E}e^{\mathcal{E}X}+D_{1}e^{\mathcal{E}X}+O\left(e^{\mathcal{E}X}\right)^{2}\right).\label{eq:disp_f_spike}
\end{equation}
Above we used $R_{\psi}=\frac{3\sqrt{3}}{2}\mathcal{J}_{\psi}\:,\:\Delta\delta=\Delta-\frac{3}{2}\mathcal{E}-\frac{1}{2}\left(\mathcal{J}_{\phi}+\mathcal{J}_{\psi}\right)\gamma$. 

The expressions for $C_{1},D_{1}$ can be found in the appendix (\ref{eq:spike_C1}),(\ref{eq:spike_D1}).
The exponential factor $X$ is of the form 
\begin{align*}
X & =-\sin^{2}\left(R_{\psi}\right)\frac{\mathcal{J}_{\phi}9\sqrt{12-81\mathcal{J}_{\phi}^{2}-36\sqrt{3}\mathcal{J}_{\phi}\cot\left(R_{\psi}\right)}}{\left(2+27\mathcal{J}_{\phi}^{2}\right)\sin^{2}\left(R_{\psi}\right)+3\mathcal{J}_{\phi}\left(9\mathcal{J}_{\phi}+2\sqrt{3}\sin\left(2R_{\psi}\right)\right)}<0.
\end{align*}

Since in this case $\mathcal{J}_{\phi}<0$, negativity of the exponent
is better recognizable from its equivalent form 
\[
X=\frac{6cd\mathcal{J}_{\phi}}{\left(c^{2}-d^{2}\right)\left(3-2\sqrt{3}\mathcal{J}_{\phi}\cot\left(R_{\psi}\right)\right)},
\]
and we have again exponentially suppressed contribution.

\section{Conclusion}

In this paper we calculated the leading finite size corrections to
the dispersion relations of giant magnon and single spike living on
 the $\gamma$-deformed Sasaki-Einstein manifold $T^{1,1}$. For
the special case $\tilde{\gamma}=0$ we get the finite size corrections
to the undeformed case discussed in \cite{Benvenuti2009}. The result
in this case has the same structure, with the difference in the $R_{\Delta}\:,\:\Delta\delta$
parameters for the magnon and the spike case respectively. An interesting
property  of the results we obtained (\ref{eq:disp_f_mag}),(\ref{eq:disp_f_spike})
is the leading $\sim\mathcal{E}e^{\mathcal{E}X}$ and the subleading
$\sim e^{\mathcal{E}X}$ part of the first order  correction. Our
results differs from the finite size correction for giant magnons
in the conformal gauge on $\mathbb{R}\times S^{2}$  discussed in
\cite{Arutyunov2007a}, which is natural from the reduced symmetry
point of view. The authors of \cite{Arutyunov2007a} considered finite
size corrections of a one spin giant magnon, whereas we are discussing
a two spin case. If we even restrict ourselves to the one spin case,
which is a nontrivial limit of the results%
\footnote{One must carefully take the limit $\omega_{\psi}\rightarrow0$ and
not just turn off the charge $\mathcal{J}_{\psi}$. %
}, we will obtain a correction for the consistent subsector of the
$AdS_{5}\times T^{1,1}$ which is quite different space than the maximally
supersymmetric $AdS_{5}\times S^{5}$. We restricted ourselves to
the first order correction since higher orders behave extremely complicated.
In principle however, one can carefully repeat the same steps and
find corrections of higher order. 

\appendix

\section{Solution to the integrals and expansions}

Using the following convenient notation $k=\frac{\left(u_{0}-u_{1}\right)\left(u_{3}-u_{2}\right)}{\left(u_{1}-u_{2}\right)\left(u_{0}-u_{3}\right)},m=\frac{u_{3}-u_{2}}{u_{1}-u_{2}}$
we obtain the explicit solutions to the integrals $I_{i}$ : 
\begin{align}
I_{1}= & -\frac{2\mathbb{K}(k)}{\sqrt{\left(u_{1}-u_{2}\right)\left(u_{0}-u_{3}\right)}}\nonumber \\
I_{2}= & -2\sqrt{\frac{1}{\left(u_{1}-u_{2}\right)\left(u_{0}-u_{3}\right)}}\left(u_{1}\mathbb{K}(k)+\left(u_{3}-u_{1}\right)\mathbb{PI}(m,k)\right)\label{eq:explicit_integrals}\\
I_{3}= & \frac{1}{\sqrt{\left(u_{1}-u_{2}\right)\left(u_{0}-u_{3}\right)}}\Bigl(-\left(u_{1}\left(u_{1}+u_{2}\right)+\left(u_{1}-u_{2}\right)u_{3}\right)\mathbb{K}(k)-\left(u_{1}-u_{2}\right)\left(u_{0}-u_{3}\right)\mathbb{E}(k)\nonumber \\
 & +\left(u_{1}-u_{3}\right)\left(u_{0}+u_{1}+u_{2}+u_{3}\right)\mathbb{PI}(m,k)\Bigr)\nonumber \\
I_{4}= & -\frac{2\sqrt{\frac{u_{1}-u_{2}}{u_{0}-u_{3}}}\left(\left(u_{3}-1\right)\mathbb{K}(k)+\left(u_{1}-u_{3}\right)\mathbb{PI}\left(\frac{1-u_{1}}{1-u_{3}}m,k\right)\right)}{\left(u_{1}-1\right)\left(u_{1}-u_{2}\right)\left(u_{3}-1\right)}\nonumber 
\end{align}

We use the following expansions for small $\epsilon$ : 
\begin{align}
\mathbb{K}\left(1-\epsilon\right)= & \left(2\log\left(2\right)-\frac{\log\left(\epsilon\right)}{2}\right)+\frac{1}{8}\left(-\log\left(\epsilon\right)+4\log\left(2\right)-2\right)\epsilon-\frac{21\epsilon^{2}}{128}-\frac{185\epsilon^{3}}{1536}+O\left(\epsilon^{4}\right)\nonumber \\
\mathbb{E}\left(1-\epsilon\right)= & 1+\frac{1}{4}\left(-\log\left(\epsilon\right)+2\log\left(4\right)-1\right)\epsilon+\frac{1}{64}\left(-6\log\left(\epsilon\right)+24\log\left(2\right)-13\right)\epsilon^{2}+\nonumber \\
 & \frac{3}{256}\left(-5\log\left(\epsilon\right)+20\log\left(2\right)-12\right)\epsilon^{3}+O\left(\epsilon^{4}\right)\label{eq:ellexpansion}\\
\mathbb{PI}\left(m,1-\epsilon\right)= & \frac{2\sqrt{m}\text{arctanh}\left(\sqrt{m}\right)+\log\left(\frac{\epsilon}{16}\right)}{2m-2}+\frac{\left(-4\sqrt{m}\text{arctanh}\left(\sqrt{m}\right)+\left(m+1\right)\log\left(\frac{16}{\epsilon}\right)-2\right)\epsilon}{8\left(m-1\right)^{2}}\nonumber \\
 & +\left(\left(12-5m\right)m+3\left(m-6\right)\log\left(\frac{16}{\epsilon}\right)m+48\text{arctanh}\left(\sqrt{m}\right)\sqrt{m}+\right.\nonumber \\
 & \left.9\log\left(\epsilon\right)-36\log\left(2\right)+21\right)\frac{\epsilon^{2}}{128\left(m-1\right)^{3}}+O\left(\epsilon^{4}\right)
\end{align}

\section{Results and useful relations }

\subsection{The giant magnon case}

Using the convenient notation $\left\{ X1=\sqrt{u0_{11}u0_{3}}\;,\; X2=\sqrt{\frac{u0_{3}}{u0_{11}}}\right\} $
we write the corrections $u1_{11,3}$ as
\begin{multline}
\left(c^{2}-d^{2}\right)^{2}u1_{11}=\frac{\mathcal{E}c^{2}d^{2}\left(-X1X2^{2}+X1-2X2\right)}{4\sqrt{3}X2^{2}(X1X2-1)}-\frac{c^{4}}{2}+c^{2}d^{2}-\frac{d^{4}}{2}-\\
\frac{d^{4}\mathcal{E}\left(-16X1X2^{4}+2X1^{2}\left(X2^{4}+8X2^{2}+3\right)X2+X1^{3}\left(X2^{6}-13X2^{4}-13X2^{2}+1\right)-32X2^{3}\right)}{48\sqrt{3}X2^{3}(X1+X2)(X1X2-1)}\\
\left(c^{2}-d^{2}\right)^{2}u1_{3}=\frac{c^{2}d^{2}\left(2X2\left(\mathcal{E}X2-2\sqrt{3}\right)+X1\left(\mathcal{E}X2\left(X2^{2}-1\right)-4\sqrt{3}\right)\right)}{4\sqrt{3}(X1+X2)}+\frac{c^{4}}{2}+\frac{d^{4}}{2}+\\
\frac{d^{4}\mathcal{E}\left(X1^{3}\left(X2^{6}-13X2^{4}-13X2^{2}+1\right)-2X1^{2}\left(3X2^{5}+8X2^{3}+X2\right)-16X1X2^{2}+32X2^{3}\right)}{48\sqrt{3}X2(X1+X2)(X1X2-1)}\label{eq:u1_corr_magnon}
\end{multline}

The correction term $\Delta_{\epsilon}=\mathcal{E}\Delta_{\mathcal{E}}+\Delta_{1}$
in (\ref{eq:disp_m_1}) consists of
\begin{align}
\Delta_{\mathcal{E}}= & \frac{d^{3}Z\sqrt{u0_{11}u0_{3}}}{324\left(\left(u0_{11}+1\right)\left(1-u0_{3}\right)\right){}^{3/2}\left(u0_{11}{}^{2}+2\left(5u0_{3}-2\right)u0_{11}+\left(u0_{3}+2\right){}^{2}\right)}\nonumber \\
 & \left(u0_{11}{}^{4}+\left(8-12u0_{3}\right)u0_{11}{}^{3}+\left(-26u0_{3}{}^{2}+8u0_{3}-36\right)u0_{11}{}^{2}-\right.\nonumber \\
 & \left.4\left(3u0_{3}{}^{3}+2u0_{3}{}^{2}+26u0_{3}-4\right)u0_{11}+\left(u0_{3}+2\right){}^{2}\left(u0_{3}{}^{2}-12u0_{3}+8\right)\right)\nonumber \\
\Delta_{1}= & \frac{1}{324\left(u0_{11}+1\right){}^{2}\left(u0_{3}-1\right)}\label{eq:Delta_mag}\\
 & \left(2Z\left(u0_{11}+1\right)\left(c^{3}\left(u0_{11}\left(9-6u0_{3}\right)-9u0_{3}+12\right)-\right.\right.\nonumber \\
 & \left.cd^{2}\left(-11u0_{3}+u0_{11}\left(2u0_{3}+11\right)+20\right)+4\sqrt{3}d^{3}\sqrt{-\left(u0_{11}+1\right)\left(u0_{3}-1\right)}\right)+\nonumber \\
 & \left.\frac{\sqrt{3}d(c^{2}-d^{2})\left(\left(u0_{11}-u0_{3}+5\right)\left(u0_{11}u0_{3}\right){}^{3/2}+4\sqrt{u0_{11}u0_{3}{}^{3}}-\sqrt{u0_{11}u0_{3}{}^{5}}\right)}{u0_{3}}\right)\nonumber \\
Z= & \arctan\left(\sqrt{\frac{\left(u0_{11}+1\right)u0_{3}}{u0_{11}-u0_{11}u0_{3}}}\right).\nonumber 
\end{align}

The coefficients of the leading $\left(C_{1}\right)$ and the subleading
$\left(D_{1}\right)$ part of the correction take the form: 
\begin{align}
C_{1}= & 8\sqrt{3}R_{\Delta}\csc\left(R_{\Delta}\right)R_{\phi}^{2}\sin^{3}\left(R_{\psi}\right)\left(-2R_{\phi}\cos\left(R_{\psi}\right)-\left(1-R_{\phi}^{2}\right)\sin\left(R_{\psi}\right)\right)\left(3-21R_{\phi}^{2}-4R_{\phi}^{4}+\right.\nonumber \\
 & \left.\left(-4+22R_{\phi}^{2}+8R_{\phi}^{4}\right)\cos\left(2R_{\psi}\right)+\left(1-R_{\phi}^{2}\right)\cos\left(4R_{\psi}\right)+8R_{\phi}\cos\left(R_{\psi}\right)\sin^{3}\left(R_{\psi}\right)+8R_{\phi}^{3}\sin\left(2R_{\psi}\right)\right)\left/\right.\nonumber \\
 & \left(\left(-1-4R_{\phi}^{2}+\left(1+2R_{\phi}^{2}\right)\cos\left(2R_{\psi}\right)+2R_{\phi}\sin\left(2R_{\psi}\right)\right)^{2}\right.\label{eq:magnon_C1}\\
 & \left.\left(R_{\phi}-3R_{\phi}\cos\left(2R_{\psi}\right)+\left(-2+R_{\phi}^{2}\right)\sin\left(2R_{\psi}\right)\right)\right)\nonumber 
\end{align}
 
\begin{align}
D_{1}= & \frac{4R_{\phi}\csc\left(R_{\Delta}\right)\sin^{2}\left(R_{\psi}\right)}{-4Y\left(R_{\phi}\cos\left(R_{\psi}\right)+2\sin\left(R_{\psi}\right)\right)\left(2R_{\phi}\sin\left(2R_{\psi}\right)+\left(2R_{\phi}{}^{2}+1\right)\cos\left(2R_{\psi}\right)-4R_{\phi}{}^{2}-1\right)}\nonumber \\
 & \left(R_{\phi}\sin\left(R_{\psi}\right)-\cos\left(R_{\psi}\right)\right)^{-1}\left(R_{\phi}{}^{2}\sin\left(2R_{\psi}\right)\left(-22YR_{\Delta}R_{\phi}-8R_{\phi}{}^{4}+38R_{\phi}{}^{2}+2\right)+\right.\label{eq:magnon_D1}\\
 & R_{\phi}{}^{2}\sin\left(4R_{\psi}\right)\left(2R_{\phi}^{4}-2YR_{\Delta}R_{\phi}^{3}-17R_{\phi}^{2}+7YR_{\Delta}R_{\phi}-1\right)+\nonumber \\
 & 4\cos\left(2R_{\psi}\right)\left(R_{\phi}\left(R_{\phi}\left(R_{\phi}\left(-5YR_{\Delta}R_{\phi}+8R_{\phi}{}^{2}-4\right)+2YR_{\Delta}\right)-2\right)-2YR_{\Delta}\right)+\nonumber \\
 & \cos\left(4R_{\psi}\right)\left(YR_{\Delta}\left(8R_{\phi}{}^{4}+R_{\phi}{}^{2}+2\right)+2R_{\phi}\left(-5R_{\phi}{}^{4}+5R_{\phi}{}^{2}+1\right)\right)+\nonumber \\
 & \left.36YR_{\Delta}R_{\phi}{}^{4}-9YR_{\Delta}R_{\phi}{}^{2}+6YR_{\Delta}-6R_{\phi}{}^{5}+6R_{\phi}{}^{3}+6R_{\phi}\right)\nonumber 
\end{align}
with $Y=\sqrt{2R_{\phi}\cot\left(R_{\psi}\right)-R_{\phi}{}^{2}+1}$.

\subsection{The single spike case}

We find the the following form for $u1_{11,3}$ 

\begin{align}
u1_{11}= & -\frac{cd\mathcal{E}u0_{11}}{2\sqrt{3}(c^{2}-d^{2})^{2}\left(u0_{11}-u0_{3}+4\right){}^{2}\sqrt{u0_{11}u0_{3}}}\Bigl(c^{2}\left(-u0_{3}{}^{2}+u0_{11}\left(u0_{11}+2\right)-8\right)\nonumber \\
 & \left(u0_{11}-u0_{3}+4\right)+6d^{2}\left(-2u0_{11}{}^{2}+\left(u0_{3}+2\right)u0_{11}+\left(u0_{3}-2\right)u0_{3}+4\right)\Bigr)-\frac{1}{2}\nonumber \\
u1_{3}= & \frac{cd\mathcal{E}u0_{3}}{2\sqrt{3}(c^{2}-d^{2})^{2}\left(u0_{11}-u0_{3}+4\right){}^{2}\sqrt{u0_{11}u0_{3}}}\Bigl(c^{2}\left(u0_{11}{}^{2}-u0_{3}{}^{2}+2u0_{3}+8\right)\nonumber \\
 & \left(u0_{11}-u0_{3}+4\right)-6d^{2}\left(u0_{11}-u0_{3}\right)\left(u0_{11}+2u0_{3}+2\right)-24d^{2}\Bigr)+\frac{1}{2}\label{eq:u1_corr_spike}
\end{align}

The parameters $\Delta_{\mathcal{E}}\,,\,\Delta_{1}$ read of 
\begin{align}
\Delta_{\mathcal{E}}= & \frac{-4d^{3}Z\sqrt{u0_{11}u0_{3}}}{3\sqrt{3}\left(u0_{11}-u0_{3}+4\right){}^{4}\left(u0_{11}{}^{2}+2\left(5u0_{3}-2\right)u0_{11}+\left(u0_{3}+2\right){}^{2}\right)}\nonumber \\
 & \left(\left(4u0_{3}-3\right)u0_{11}{}^{4}+\left(4u0_{3}{}^{2}+4u0_{3}+2\right)u0_{11}{}^{3}+\right.\nonumber \\
 & 2\left(-2u0_{3}{}^{3}+7u0_{3}{}^{2}+u0_{3}+12\right)u0_{11}{}^{2}-2\left(2u0_{3}{}^{4}-2u0_{3}{}^{3}+u0_{3}{}^{2}-40u0_{3}+12\right)u0_{11}-\nonumber \\
 & \left.\left(u0_{3}+2\right){}^{2}\left(3u0_{3}{}^{2}-10u0_{3}+4\right)\right)\nonumber \\
\Delta_{1}= & \frac{1}{27\left(u0_{11}-u0_{3}+4\right){}^{2}}\left(\sqrt{3}d\left(d^{2}-c^{2}\right)\sqrt{u0_{11}u0_{3}}\left(u0_{11}-u0_{3}+4\right)+\right.\label{eq:Delta_spike}\\
 & \frac{2Z}{c}\left(2c^{4}\left(u0_{11}-u0_{3}+4\right)-c^{2}d^{2}\left(-11u0_{3}+u0_{11}\left(2u0_{3}+11\right)+20\right)+\right.\nonumber \\
 & \left.\left.3d^{4}\left(u0_{11}\left(3-2u0_{3}\right)-3u0_{3}+4\right)\right)\right)\nonumber \\
Z= & \arctan\left(\sqrt{\frac{\left(u0_{11}+1\right)u0_{3}}{u0_{11}-u0_{11}u0_{3}}}\right).\nonumber 
\end{align}

The correction coefficients $C_{1},D_{1}$ for the single spike dispersion
relation are of the form 
\begin{align}
C_{1} & =162\mathcal{J}_{\phi}^{2}\Delta\delta\csc\left(\Delta\delta\right)\sqrt{4-27\mathcal{J}_{\phi}^{2}-12\sqrt{3}\mathcal{J}_{\phi}\cot\left(R_{\psi}\right)}\sin^{4}\left(R_{\psi}\right)\left(12-810\mathcal{J}_{\phi}^{2}+1458\mathcal{J}_{\phi}^{4}+\right.\nonumber \\
 & \left(-16+864\mathcal{J}_{\phi}^{2}+2916\mathcal{J}_{\phi}^{4}\right)\cos\left(2R_{\psi}\right)+\sqrt{3}\left(-36\mathcal{J}_{\phi}+162\mathcal{J}_{\phi}^{3}+2187\mathcal{J}_{\phi}^{5}\right)\sin\left(2R_{\psi}\right)\nonumber \\
 & \left.+\left(4-54\mathcal{J}_{\phi}^{2}\right)\cos\left(4R_{\psi}\right)+18\sqrt{3}\mathcal{J}_{\phi}\sin\left(4R_{\psi}\right)\right)\left/\biggl(\left(3\sqrt{3}\mathcal{J}_{\phi}\cos\left(R_{\psi}\right)-4\sin\left(R_{\psi}\right)\right)\right.\label{eq:spike_C1}\\
 & \left(2\cos\left(R_{\psi}\right)+3\sqrt{3}\mathcal{J}_{\phi}\sin\left(R_{\psi}\right)\right)\left(2+54\mathcal{J}_{\phi}^{2}-\left(2+27\mathcal{J}_{\phi}^{2}\right)\cos\left(2R_{\psi}\right)+6\sqrt{3}\mathcal{J}_{\phi}\sin\left(2R_{\psi}\right)\right)^{2}\biggr)\nonumber 
\end{align}

\begin{align}
D_{1}= & \frac{-3\mathcal{J}_{\phi}\csc(\Delta\delta)\sin^{2}\left(R_{\psi}\right)\left(3\sqrt{3}\mathcal{J}_{\phi}\sin\left(R_{\psi}\right)+2\cos\left(R_{\psi}\right)\right)^{-1}}{4\left(3\sqrt{3}\mathcal{J}_{\phi}\cos\left(R_{\psi}\right)-4\sin\left(R_{\psi}\right)\right)\left(6\sqrt{3}\mathcal{J}_{\phi}\sin\left(2R_{\psi}\right)-\left(27\mathcal{J}_{\phi}{}^{2}+2\right)\cos\left(2R_{\psi}\right)+54\mathcal{J}_{\phi}{}^{2}+2\right)}\nonumber \\
 & \biggl(8\left(\sqrt{3}\Delta\delta\left(-3645\mathcal{J}_{\phi}{}^{4}+216\mathcal{J}_{\phi}{}^{2}-32\right)+9Y\mathcal{J}_{\phi}\left(81\mathcal{J}_{\phi}{}^{2}+4\right)\right)\cos\left(2R_{\psi}\right)+\nonumber \\
 & 2\left(4\sqrt{3}\Delta\delta\left(27\left(54\mathcal{J}_{\phi}{}^{4}+\mathcal{J}_{\phi}{}^{2}\right)+8\right)-9Y\mathcal{J}_{\phi}\left(81\mathcal{J}_{\phi}{}^{2}+4\right)\right)\cos\left(4R_{\psi}\right)+6\Bigl(27\mathcal{J}_{\phi}{}^{2}\sin\left(2R_{\psi}\right)\nonumber \\
 & \left(\sqrt{3}Y\left(\left(2-9\mathcal{J}_{\phi}{}^{2}\right)\cos\left(2R_{\psi}\right)+18\mathcal{J}_{\phi}{}^{2}-2\right)+6\Delta\delta\mathcal{J}_{\phi}\left(\left(27\mathcal{J}_{\phi}{}^{2}-14\right)\cos\left(2R_{\psi}\right)+22\right)\right)-\nonumber \\
 & 9Y\left(81\mathcal{J}_{\phi}{}^{2}+4\right)\mathcal{J}_{\phi}+4\sqrt{3}\Delta\delta\left(2187\mathcal{J}_{\phi}{}^{4}-81\mathcal{J}_{\phi}{}^{2}+8\right)\Bigr)\biggr)\label{eq:spike_D1}
\end{align}
where we used the notation $Y=\sqrt{-12\sqrt{3}\mathcal{J}_{\phi}\cot\left(R_{\psi}\right)-27\mathcal{J}_{\phi}{}^{2}+4}$.\bibliographystyle{ieeetr}
\bibliography{fs}

\end{document}